\hoffset=1truecm
\hsize=15truecm

\vsize=22truecm 

\baselineskip=13pt
\hfuzz=60pt
\def\const{{\rm const}\,}
\def \prf{\vskip.25truein\noindent{\bf Proof\ }:\ \ \ }
\def\endproof{\hfill\vrule height .6em 
width .6em depth 0pt\goodbreak\vskip.25in }
\def\cA{{\cal A}}\def\cG{{\cal G}}
\def\sfrac#1#2{{\textstyle{#1\over#2}}}
\def\half{\sfrac{1}{2}} 
\def\<{\left<}
\def\>{\right>}

\def\de{\delta}
\def\ep{\epsilon}

\def\th{\theta}
\def\ka{\kappa}
\def\la{\lambda}

\def\si{\sigma}
\def\ta{\tau}

\def\ph{\phi}
\def\ch{\chi}
\def\ps{\psi}

\def\De{\Delta}

\def\La{\Lambda}

\def\norm{\Vert}
\def\D{{\cal D}}
\def\x{{\bf x }}
\def\y{{\bar {\bf x }}}

\def\cD{{\cal D}}

\def\bbbr{{\rm I\!R}} 
\def\k{{\bf k}}
\def\p{{\bf p}}
\def\q{{\bf q}}
\def\optbar#1{\vbox{\ialign{##\crcr\hfil${\scriptscriptstyle(}\mkern -1mu
         \vrule height 1.2pt width 3pt depth -.8pt
         {\scriptscriptstyle)}$\hfil\crcr
          \noalign{\kern-1pt\nointerlineskip}$\hfil
\displaystyle{#1}\hfil$\crcr}}}
\centerline{\bf A Single Scale
Infinite Volume Expansion}
\centerline{\bf for Three-Dimensional Many Fermion 
Green's Functions}
\centerline{Jacques Magnen,} 
\centerline{Vincent Rivasseau}
\centerline{Centre de Physique Th\'eorique, CNRS UPR 14}
\centerline{Ecole Polytechnique, F-91128 Palaiseau Cedex, FRANCE}
\vskip2truecm

\noindent
\underbar{Abstract}\ \ \ 
In [FMRT1] we introduced a multiscale expansion 
for many Fermion systems in two space dimensions
based on a so-called sector decomposition. 
In this paper a completely different
expansion is introduced to treat the more difficult case of three
(or more) space dimensions: it is based on an auxiliary scale decomposition
and the use of the Hadamard inequality.
We prove that the perturbative expansion for a {\it single scale}
model has a convergence radius {\it independent of the scale}.
This is a typical result, already proved in the two dimensional
case in [FMRT1], which we cannot obtain 
in three dimensions by naive extrapolation of the sector method. 
Although we do not treat in this
paper the full (multiscale) system, we hope 
this new method to be a significant step towards the
rigorous construction of the BCS theory of superconductivity in three
space dimensions.

\vskip 2cm

\noindent{\bf I Introduction}
\vskip.25truein
 
In this paper we consider many Fermion systems formally characterized
by the effective potential 
$$ {\cal G}(\psi^e,\bar \psi^e)
=\log\,{1\over Z}\int e^{-{\cal
V}(\psi+\psi^e,\bar\psi+\bar\psi^e)} d\mu_{C}(\psi,\bar\psi) \eqno({\rm I.0}) 
$$ 
for
the external fields $\psi^e,\bar\psi^e$. Here, $d\mu_C(\psi,\bar\psi)$
is the Fermionic Gaussian measure in the Grassmann variables
$\{\psi(\xi),\bar\psi(\xi)\ |\ \xi\in
\bbbr\times\bbbr^d\times\{\uparrow,\downarrow\}\}$ with propagator $$
C(\xi,\bar\xi)=\delta_{\sigma,\bar\sigma}\int
{d^{d+1}p\over(2\pi)^{d+1}} {e^{i<p,\xi-\bar\xi>_-}
\over ip_0-e(\p)}
\eqno({\rm I.1}) 
$$where 
$$
 <p,\xi>_-=\p\cdot\x-p_0\tau \eqno({\rm I.2})
$$ 
and 
$$
 e(\p)={\p^2\over2m}-\mu.  \eqno({\rm I.3})
$$ 
The variable $\xi=(\tau,\x,\sigma)$ consists of time,
space and spin components and the $(d+1)$-momentum $p=(p_0,\p)$. The 
interaction is given by 
$$ 
{\cal V}={\la\over 2}
\int \prod_{i=1}^4 d\xi_i\ \  
V\!\left(\xi_1,\xi_2,\xi_3,\xi_4\right)
\bar\psi(\xi_1)\bar\psi(\xi_2)\psi(\xi_4)\psi(\xi_3)\eqno({\rm I.4})
$$ 
where $\la$ is the coupling constant,
the kernel $V\!\left(\xi_1,\xi_2,\xi_3,\xi_4\right)$ is
translation invariant with $V\!\left(0,\xi_2,\xi_3,\xi_4\right)$
integrable and 
$$
\int d\xi=\sum_{\sigma\in\{\uparrow,\downarrow\}}
\int_\bbbr d\tau\int_{\bbbr^d} d\x.\eqno({\rm I.5})
$$ The partition function is
$$ Z=\int e^{-{\cal
V}(\psi,\bar\psi)}d\mu_{C}(\psi,\bar\psi) 
\eqno({\rm I.6})$$ so that ${\cal
G}(0,0)=0.$

The fact that $\cal V$ is non-local (although short range) is the
source of a few inessential complications. 
Therefore in this paper we restrict ourselves
to the case of a $\de$ function
to focus the reader's attention on essential aspects.
Therefore from now on, we will consider
$$ 
{\cal V}_{\La}={\la\over 2}
\int_{\La}  d\x d\ta  \ 
\bar\psi_{\uparrow}(\x, \ta)\bar\psi_{\downarrow}
(\x, \ta)\psi_{\uparrow}(\x, \ta)\psi_{\downarrow}(\x, \ta)
\eqno({I.4.b})
$$

The Euclidean Green's functions with $2p$ points $$
G_p(\xi_1,\bar\xi_1,\dots,\xi_p,\bar\xi_p)=\prod_{i=1}^p
{\delta^2\hfill\over\delta\psi^e(\xi_i)\delta\bar\psi^e(\bar\xi_i)}{\cal
G} 
\eqno({\rm I.7})$$ generated
by the effective potential are the connected Green's functions
amputated by the free propagator. By definition, ${\cal G}$ exists
when the norm $$
\norm G_p\norm=\max_{j}\ \sup_{\xi_j}\ \int\prod_{i\ne j}d\xi_i\  
|G_p(\xi_1,\cdots,\xi_{2p})| \eqno({\rm I.8})
$$ of each of its moments, $G_p,\ p\ge
1$, is finite. Intuitively, $\norm G_p\norm$ is the supremum in
momentum space of $G_p$. In fact, the supremum in momentum space was
used as the standard norm on vertices in [FT2].

Our goal is to give a rigorous proof that the standard model
for a weakly interacting system of electrons and phonons has a
superconducting ground state at sufficiently low temperature.
Perturbation theory and, in particular, the renormalization of the two
point function was controlled in [FT1]. 
A renormalization group flow
for the four point function was defined and analyzed in [FT2]. Two
additional ingredients are required to complete this program.  First
we need an infinite volume expansion that
controls non-perturbatively
this renormalization group flow. This first ingredient, 
``constructive stability''
was provided in two space dimensions (d=2) in [FMRT1]. 
The other physically interesting case, namely $d=3$, is much harder.
In this paper we prove by a completely different method
that one of the key results of [FMRT1]
extends to $d=3$: the radius of convergence of the single
scale theory is uniformly bounded below. However up to now, we have
not been able to extend this result to the multiscale theory.
Therefore ``constructive stability'' in $d=3$
remains an open problem.

The second ingredient is the control of the reduced BCS model for
Cooper pairs via another expansion (of the $1/N$ type). One has to
prove existence of the BCS gap, of the associated spontaneous $U(1)$
symmetry breaking, and of the masslessness of the associated 
Goldstone boson. This program is well under way. For references
on this part of the program and on the general strategy see [FMRT3-5]. 

In the remainder of this paper we specialize to  $d=3$. Our
method extends however without difficulty to any $d\ge 2$
(see Appendix I).

As in [FT1,2], the model is sliced into energy regimes by decomposing
momentum space into shells around the Fermi surface. The $j^{\rm th}$
slice has covariance $$
C^{(j)}(\xi,\bar\xi)=\delta_{\sigma,\bar\sigma}\int
{d^{4}p\over(2\pi)^{4}} {e^{i<p,\xi-\bar \xi>_-}
\over ip_0-e(\p)}f_j(p),\eqno({\rm I.9})
$$ where $$ f_j(p)=f\left(M^{-2j}\left(p_0^2+e(\p)^2\right)\right) 
\eqno({\rm I.10})$$
effectively forces $|ip_0-e(\p)|\sim M^j$.  The function $f\in
C^\infty_0([1,M^{4}])$. The parameter $M$ is strictly bigger than one so
that the scales near the Fermi surface have $j$ near $-\infty$. The
model is defined in finite volume and at fixed scale by the following
lemma:
 
\medskip
\noindent{\bf Lemma I.1}{\it \ \
$$ {\cal G}_\Lambda^{(j)}(\psi^e,\bar\psi^e) =\log\,{1\over
Z_\Lambda^{(j)}}\int e^{-{\cal
V}_\Lambda(\psi+\psi^e,\bar\psi+\bar\psi^e)}
d\mu_{C^{(j)}}(\psi,\bar\psi) 
\eqno({\rm I.11})
$$ where $$ {\cal V}_\Lambda={\lambda\over 2}
\int_{\Lambda^4} \prod_{i=1}^4 d\xi_i\ \  
V\!\left(\xi_1,\xi_2,\xi_3,\xi_4\right)
\bar\psi(\xi_1)\bar\psi(\xi_2)\psi(\xi_4)\psi(\xi_3)
\eqno({\rm I.12})$$
 and 
$$ Z_\Lambda^{(j)}=\int e^{-{\cal V}_\Lambda(\psi,\bar\psi)} 
d\mu_{C^{(j)}}(\psi,\bar\psi) 
\eqno({\rm I.13})$$ 
is
analytic in $\lambda$ in a neighborhood of the origin that includes 
the disk of radius $\const\left(M^{2j}|\Lambda|\right)^{-1}.$}

\medskip
The full proof (not very difficult) is given in [FMRT1, Lemma 1]. 
Remark however
that the radius of convergence depends on volume and scale in a
patently unsatisfactory way.

Recall that $G^{(j,\Lambda)}_p$ 
is the $2p$-point Green's function generated by
${\cal G}^{(j)}_\Lambda$. By the Lemma above the Taylor series 
$$
G^{(j,\Lambda)}_p=\sum_{n=0}^\infty g_n(p,j,\Lambda)\lambda^n 
\eqno({\rm I.14})$$ 
has a
strictly positive, though possibly $j$ and $\Lambda$ dependent, radius
of convergence. The main result of this paper is

\medskip
\noindent{\bf Theorem I}{\it \ \ 
There exists a $\ \const $, independent of $j$ and 
$\Lambda$, such that $$
\norm g_n(p,j,\Lambda)\norm \le\const^{n+p}
M^{(4-2p)j}\norm V\norm^n
\eqno({\rm I.15})$$ 
where 
$$
\norm g_n(p,j,\Lambda)\norm=
\max_{k}\ \sup_{\xi_k}\ \int\prod_{i\ne k}d\xi_i\  
|g_n(p,j,\Lambda)(\xi_1,\cdots,\xi_{2p})|.  
\eqno({\rm I.16})$$ Furthermore the limits
$$ 
g_n(p,j)=\lim_{\Lambda\rightarrow \bbbr^3}g_n(p,j,\Lambda) 
\eqno({\rm I.17})$$ 
exist and the infinite volume Green's functions at scale $j$ 
$$
G^{(j)}_p=\sum_{n=0}^\infty g_n(p,j)\lambda^n 
\eqno({\rm I.18})$$ 
are analytic in
$|\lambda|<R =({\const} \norm V\norm)^{-1}$.}
\vskip .25truein

This theorem is 
analogous to the first theorem of [FMRT1], which applied only to $d=2$.
Only the dependence in the external fields is different. Remark
that this external field dependence, i.e. the factor 
$M^{(4-2p)j}$ is still not the 
desired perturbative one which
would be $M^{{1/2}(4-2p)j}$ [FT1].

A theorem such as Theorem I is usually proven with a standard cluster/Mayer
expansion. Space-time, $\bbbr^{d+1}$, is paved by cubes
$\Delta$ of side $M^{-j}$ dual to the decay rate $M^j$ of the
propagator. The decay rate is primarily determined by the thickness of
the shell in momentum space. Then one expands in coupling constants
that control the interaction between boxes. One essential prerequisite
for the convergence of such expansions is the $j$ independent estimate
$|Z(\Delta)|\le\const$. For our models, one can see in perturbation
theory that this estimate fails. For instance the vacuum graph $G_{2}$
(see Fig.I) can be computed explicitly and in $d=3$ it
leads to a second order contribution in $e^{\la^{2}M^{-j}}$
to $|Z(\Delta)|$. 

This difficulty can be traced to the fact that the Pauli exclusion principle
now permits about $M^{-(d-1)j}$ electrons to be located in $\Delta$ with
momentum restricted to the shell. For $d=1$ the Fermi
surface consists of just two points 
and consequently only $O(1)$ electrons are allowed in $\Delta$
and there is no difficulty. As $d$ grows the
Pauli exclusion principle becomes progressively weaker and the
estimate on the partition function in $\Delta$ becomes more 
and more $j$ dependent.

In [FMRT1] the solution to this difficulty was found in a decomposition
of the shell into smaller units, called sectors. This idea does not 
seem to work beyond two dimensions.
An informal discussion of why this is so is included 
for the interested reader in Appendix II. Our proof
of Theorem I relies therefore 
on a completely different idea~: we decompose the
propagator in a slice according to an auxiliary scale, and 
the Hadamard inequality is used to
control the Fermionic determinants. 
\vskip .5cm

\noindent
{\bf Acknowledgements}\ \ This paper is part of a larger program
in collaboration with J. Feldman and E. Trubowitz. We thank particularly
J. Feldman for his comments.

\medskip
\noindent{\bf II The auxiliary scale decomposition}
\vskip.25truein

The $j$-th slice propagator is
$$
C^{j}(\xi, \bar \xi) = \de_{\si\bar \si}
\int {dp_{0}d^{3} \p 
\over (2\pi)^{4}}{f_{j}(p) \over ip_{0}-e(\p)}e^{i<p
(\xi - \bar \xi)>_{-}}
$$
$$ = \de_{\si\bar \si}{1\over (2\pi)^{4}}
\int_{-\infty}^{+\infty} dp_{0} e^{ip_{0}(\bar \ta -\ta)}\int_{-1}^{+\infty}
du \ f_{j}(p_{0},u) D(p_{0},u)
 { 4\pi \sin ( \sqrt{1+u}| \x- \y|) 
\over 2|\x-\y|}
$$
$$ = \de_{\si\bar \si}{1 \over 8\pi^{3}|\x-\y|} 
\int_{-M^{j+2}}^{+M^{j+2}}\int_{-M^{j+2}}^{+M^{j+2}} 
dp_{0} du\  e^{ip_{0}(\bar \ta -\ta)}
\ f_{j}(p_{0},u) D(p_{0},u)
 \sin ( \sqrt{1+u}| \x- \y|)  
 \eqno({\rm II.1})$$
by putting $u=e(\p)$, $f_{j}(p_{0},u)\equiv f(M^{-2j}(p_{0}^{2}+u^{2}
))$, and $D(p_{0},u)= (ip_{0}-u)^{-1}$. This formula is proved by 
performing the integrals over the angular components $\th$
and $\ph$ of $\p$. In the last
line we used the fact that $f$ has its support in $[1, M^{4}]$,
and we assumed that $j \le -2$\footnote*{The first slices $j=0$ or $j=-1$ are 
trivial from the point of view of this paper.}.

We decompose further this propagator into {\it auxiliary slices}
according to the scale of $|\x - \y|$ by writing
 
$$
C^{j}(\xi, \bar \xi) = \sum_{k=j}^{0} 
C^{j,k}(\xi,\bar \xi)
\eqno({\rm II.2})$$
with:
$$C^{j,k}(\xi,\bar \xi) =-  A^{j,k} (\x,\y)
\de_{\si\bar \si} {1 \over 8\pi^{3} |\x -\y|}
 \int_{-M^{j+2}}^{+M^{j+2}} 
\int_{-M^{j+2}}^{+M^{j+2}}
dp_{0} du e^{ip_{0}(\bar \tau -\tau )}
$$
$$D(p_{0},u)
f_{j}(p_{0},u) \sin (\sqrt{1+u}|\x -\y |)  
 \eqno({\rm II.3})$$
$$
 A^{j,0} (\x,\y) = e^{-|\x -\y|^{2}}
\eqno({\rm II.4a})
$$
$$
 A^{j,k} (\x,\y) =  e^{-M^{2k}|\x -\y|^{2}} - e^{-M^{2k+2}|\x -\y|^{2}} 
\ \ {\rm for \  -1 \ge k \ge j+1}
\eqno({\rm II.4b})
$$
$$
 A^{j,j} (\x,\y) =  1- e^{-M^{2j+2}|\x -\y|^{2}} 
\eqno({\rm II.4c})
$$
These formulas 
are totally explicit. We choose them so that $C^{j,k}$ is a nice convolution of
$C^{j}$ with an explicit spatial Gaussian, since the Fourier transform of
a Gaussian is Gaussian, but other formulas are possible. We have 

\noindent{\bf Lemma II.1}{\it \ \ There exists 
a constant $K(p)$ such that, for 
$k=0,...,j$, and any integer $p$
the functions $C^{j,k}$ obey the bounds~:
$$
\big| 
C^{j,k}(\xi,\bar \xi) \big| 
\le K(p) M^{j+ k} 
(1+M^{k}|\x -\y|)^{-p} (1+M^{j}|\tau -\bar \tau|)^{-p} 
\eqno({\rm II.5a})
$$
Furthermore
$$C^{j,k}(\xi,\xi)  = 0 {\rm \ for \ } \ -1 \ge k \ge j\ ; \ \big| 
C^{j,0}(\xi,\xi) \big| 
\le M^{2j}
\eqno({\rm II.5b})
$$
}
\prf  
For $k \ne j$, we use the identity 
$$(1+iM^{j}(\tau -\bar\tau ))^{p} e^{ip_{0}(\bar \tau -\tau )}
=  (1+ M^{j}{d \over d p_{0}})^{p} e^{ip_{0}(\bar \tau -\tau )}
\eqno({\rm II.6})$$
and the obvious inequality $(1+M^{j}|\tau -\bar\tau |) \le \sqrt{2}
|1+iM^{j}(\tau -\bar\tau )|$
to write
$$(1+M^{j}|\tau -\bar\tau |)^{p} |C^{j,k}| \le
(\sqrt{2})^{p} \biggl| A^{j,k} (\x,\y)
\de_{\si\bar \si} {1 \over 8\pi^{3} |\x -\y|}
 \int_{-M^{j+2}}^{+M^{j+2}} 
\int_{-M^{j+2}}^{+M^{j+2}}
dp_{0} du 
$$
$$\biggl( (1+ M^{j}{d \over d p_{0}})^{p} e^{ip_{0}(\bar \tau -\tau )}
\biggr)
D(p_{0},u)
f_{j}(p_{0},u) \sin (\sqrt{1+u}|\x -\y |) \biggr|
$$
$$ = (\sqrt{2})^{p}\biggl| A^{j,k} (\x,\y)
\de_{\si\bar \si} {1 \over 8\pi^{3} |\x -\y|}
 \int_{-M^{j+2}}^{+M^{j+2}} 
\int_{-M^{j+2}}^{+M^{j+2}}
dp_{0} du   e^{ip_{0}(\bar \tau -\tau )}
$$
$$\biggl((1- M^{j}{d \over d p_{0}})^{p} D(p_{0},u)
f_{j}(p_{0},u) \biggr) \sin (\sqrt{1+u}|\x -\y |) \biggr|
\eqno({\rm II.7})$$
The last line is obtained by integration by parts;
boundary terms vanish
by the support properties of $f_{j}$.
Computing the action of 
partial derivatives $M^{j}{d \over d p_{0}}$ on 
$D(p_{0},u) f_{j}(p_{0}, u )$
completes the proof. Indeed by the scaling and support properties of
$f_{j}$ we obtain, up to a constant $K(p)$, the same bound as for
the initial function $D(p_{0},u) f_{j}(p_{0}, u )$. 
More precisely
for  $k\ne 0$ we bound $\sin (\sqrt{1+u}| \x -\y |)$ by 1. Using then
$$ \int_{-M^{j+2}}^{+M^{j+2}} 
\int_{-M^{j+2}}^{+M^{j+2}}
dp_{0} du  | D(p_{0},u) | \le M^{j} \ .
\eqno({\rm II.8})$$
we obtain
that 
$$(1+M^{j}|\tau -\bar\tau |)^{p} |C^{j,k}| \le K_{1}(p)  M^{j}
\biggl| { A^{j,k} (\x,\y)\over  \x -\y }\biggr| \ ,
\eqno({\rm II.9})$$
and we remark that for $k\ne j$, $k \ne 0$ and for any $p$
$$\biggl| { A^{j,k} (\x,\y)\over  \x -\y }\biggr| \le 
K_{2}(p)M^{k} (1+M^{k}|\x -\y |)^{-p} \ .
\eqno({\rm II.10})$$
(In fact the function $A^{k}$
has exponential decrease for $k\ne j$). 
In the case $k=0$, we use the bound
$$ { \sin (\sqrt{1+u}| \x -\y |) \over |\x -\y | }  \le \sqrt{2}
\eqno({\rm II.11})$$
for $|u |\le 1$ (which is true on the integration domain). We combine
this bound with the trivial estimate 
$$\biggl|  A^{j,0} (\x,\y) \biggr| \le 
K_{2}(p)  (1+ |\x -\y |)^{-p}
\eqno({\rm II.12})$$
to obtain (II.5a).

In the last case $k=j$, the spatial decay also has to 
come from integration by parts. For simplicity we assume that $p=2q$ is even 
(this obviously implies the general case as well).
We use the identity:
$$M^{2j}|\x -\y |^{2} \sin (\sqrt {1+u}|\x -\y |) = 
(2iM^{j} \sqrt{1+u} {d \over du})^{2} \sin (\sqrt {1+u}|\x -\y |)
\eqno({\rm II.13})$$

Therefore we have 
$$(1+M^{j}|\x -\y |)^{p}(1+M^{j}|\tau -\bar\tau |)^{p} |C^{j,j}| \le 2^{p} 
\biggl|(1+M^{2j}|\x -\y |^{2})
^{q}(1+iM^{j}(\tau -\bar\tau ))^{p} C^{j,j} \biggr|
$$
$$ = 2^{p} \biggl| A^{j,j} (\x,\y)
\de_{\si\bar \si} {1 \over 8\pi^{3} |\x -\y|}
 \int_{-M^{j+2}}^{+M^{j+2}} 
\int_{-M^{j+2}}^{+M^{j+2}}
dp_{0} du \biggl( (1+ M^{j}{d \over d p_{0}})^{p} e^{ip_{0}(\bar \tau -\tau )}
\biggr)
$$
$$D(p_{0},u)
f_{j}(p_{0},u) 
\biggl(\bigl(1+ (2 i M^{j}\sqrt{1+u} {d \over du})^{2}\bigr)^{q}
\sin (\sqrt{1+u}|\x -\y |) \biggr) 
\biggr|
\eqno({\rm II.14})$$
Then we integrate by parts, both on $p_{0}$ and $u$. 
The boundary terms vanish again
by the support properties of $f_{j}$, hence
partial derivatives $(1-M^{j}{d \over d p_{0}})$ 
act as before on $D(p_{0},u)
f_{j}(p_{0}, u )$, and operators $2 i M^{j}\sqrt{1+u} {d \over du}$
change into transposed operators  
$(-2 i M^{j} {d \over du} ) \sqrt{1+u}\; \cdot \; $, 
also acting  on $D(p_{0},u)
f_{j}(p_{0},u)$. This action, by Leibniz rule, 
is slightly complicated to write down. But clearly,
again by the support and scaling properties of $f_{j}$,
we get up to a constant
$K(p)$ the same estimate than for the initial $p_{0}$ and $u$ integral,
namely $M^{j}$. We bound $|\sin (\sqrt{1+u}|\x -\y |)|$ by one, and
the last factor $M^{k}=M^{j}$ for $k=j$
comes from the easy estimate on (II.4c): 
$$\biggl| { A^{j,j} (\x,\y)\over  \x -\y }\biggr| \le M^{j} 
\eqno({\rm II.15})$$

Finally (II.5b) is easy and left to the reader. (For the case $k=0$,
use the fact that $\int dp_{0}du D(p_0 ,u ) f_j (p_0 , u) =0$.)
\endproof

Corresponding to this division of the covariance, there is an associated
orthogonal decomposition of the fields $\ps^{j}$ and $\bar \ps^{j}$
distributed with covariance $C^{j}$ as sums of fields 
$\ps^{j,k}$ and $\bar \ps^{j,k}$
distributed with covariance $C^{j,k}$:
$$
\psi^{j} (\xi) = \oplus_{k=j}^{0} 
\psi^{j,k} (\xi)
\ ; \ 
\bar \psi^{j}  (\bar \xi) = \oplus_{k=j}^{0} 
\bar \psi^{j,k} (\bar \xi)
\eqno({\rm II.16})
$$

For each scale $k$ we shall perform in the next section a cluster expansion
according to a lattice $\D_{k}$ of tubes of spatial side $M^{-k}$
and of time side $M^{-j}$. 
The power counting with respect to this auxiliary scale
is superrenormalizable and identical to the one
of the ultraviolet limit of the $\ph^{4}_{3}$. 
To understand this fact, we can compute the weight 
of a connected graph with $n$ vertices and $l$ internal lines
at subscale $(j,k)$ with
$k>j$ and $ e= 2p = 4n -2l$ external legs.
Using the same norm as in Theorem I of section 1, we fix a particular
vertex and integrate over the others.
The usual estimate, selecting a tree in the graph [R], gives
$M^{-(j+3 k) (n-1)}$ for  integration of the vertices
times $M^{(j+k)l}$ for the power counting of the lines.
This is $M^{(4-2p)j} M^{(n-3+p)(j-k)}$. The factor $M^{(4-2p)j}$
is that of (I.15) in Theorem I.
The sum over $k$ is 
convergent provided $n+p -3 >0$, which selects exactly the 
same graphs as the condition of ultraviolet
convergence in the $\ph^{4}_{3}$ theory. This is to our knowledge the
first example of a Fermionic model with this kind of power counting.

\vskip.3truecm
\vskip.3truecm
\centerline{Figure 1: The divergent graphs in d=3}
\medskip


We conclude that
as far as renormalization
is concerned, the {\it single slice} 
model treated in this paper is 
similar in  difficulty to the old ultraviolet construction of 
$\ph^{4}_{3}$ by now well established [GJ]. 
Vacuum graphs are
divergent up to three vertices hence the graphs $G_{1}$
$G_{2}$, $G'_{2}$ and $G_{3}$, $G'_{3}$,
$G'_{3}$... of Fig. 1, are divergent and the
two point subgraphs are 
divergent up to two vertices, hence the graphs $B_{1}$
and $B_{2}$, $B'_{2}$ of Fig. 1, are a priori divergent.
But by (II.5b) the ``tadpole'' is convergent, hence
$G_{1}$, $B_{1}$ and all the ``primed'' graphs in
fact converge. The second order vacuum graph
$G_{2}$ is linearly divergent and has to be cancelled
explicitly by going to intensive quantities such
as the pressure or normalized Schwinger functions.
Finally the  graph $B_{2}$ is apparently logarithmically divergent
if we replace the propagators by the bound (II.5a). But in
fact it is convergent because of the oscillations of the
$\sin^{3}$ function, which are lost in the right hand side of (II.5a).
To exploit this fact, one must
renormalize this graph, hence make an explicit computation
for it. Had this graph not been convergent,
we would have been able only to prove
a bound in $|j|^{-1/2}$ on the radius of convergence in Theorem I.

\medskip
\noindent{\bf III The expansion}
\vskip.25truein
Since this problem is superrenormalizable, we can treat
the few divergent graphs by introducing explicit counterterms,
in the manner of [GJ].
This is no problem for the vacuum
graphs since they will quotient out anyway
in the normalized functions. But for
the two point subgraphs $B_{1}$ and $B_{2}$, we have to compensate
for these counterterms by changing the propagator:

\medskip
\noindent{\bf Lemma III.1}{\it\ \  The
infinite volume connected Green's functions defined by
$$
G_p(\xi_1,\bar\xi_1,\dots,\xi_p,\bar\xi_p)=\prod_{i=1}^p
{\delta^2\hfill\over\delta\psi^e(\xi_i)\delta\bar\psi^e(\bar\xi_i)}{\cal
G} 
\eqno({\rm III.1})$$ 
$$ {\cal G}(\psi^e,\bar \psi^e)
=\log\,{1\over Z}\int e^{-{\cal
V}(\psi+\psi^e,\bar\psi+\bar\psi^e)} d\mu_{C}(\psi,\bar\psi)  
\eqno({\rm III.2})$$ 
are identical to those of the theory
$$ \bar { \cal G}(\psi^e,\bar \psi^e)
=\log\,{1\over \bar Z}\int e^{-{\cal I}} d\mu_{\bar C}(\psi,\bar\psi) 
\eqno({\rm III.3})$$
where 
$$ \bar C (p)= \sum_{n=0}^{\infty}
C (p)\bigl(\hat B(\la, p) C(p)\bigr)^{n}
\eqno({\rm III.4})$$
$$
{\cal I}= {\cal V}(\psi + \psi_e , \bar \psi + \bar \psi _e ) 
+ {\cal M} + {\cal W}
\eqno({\rm III.5})$$
$$ {\cal M} = \la \int d\xi B_{1} \bar\psi(\xi)\ps (\xi) + \la^{2}
\int d\xi_1 d\xi_{2}\ \bar\psi(\xi_1) 
B_{2}(\xi_{1},\xi_{2})\psi(\xi_2)  
\eqno({\rm III.6})$$
$$ {\cal W} = \int  ( a\la + b \la^{2} + c\la^{3} )
\eqno({\rm III.7})$$
where $\hat B(\la, p)$ is the Fourier transform of the translation 
invariant kernel $\la B_{1}\de( \xi_{1},\xi_{2})  
+ \la^{2} B_{2}( \xi_{1},\xi_{2})$.
}

\prf In the perturbation expansion
the counterterm ${\cal W}$ cancels out, and
the formal power series expansion of the propagator
$\bar C$ exactly cancels against the insertions of the
mass counterterm ${\cal M}$ to give back the original propagator
$C$. But by Lemma I.1, the perturbation expansion really
defines the theory. This proves that the two theories
are also identical at the constructive level, as 
defined by the same analytic germ. \endproof

We apply this lemma with $a$ $b$ and $c$ in ${\cal W}$ the amplitudes
of the divergent vacuum graphs with one, two and three vertices
in Fig. 1, and 
$$  {\cal M} =\la \int d\xi B_{1} \bar\psi(\xi)\ps (\xi) + \la^{2}
\int d\xi_1 d\xi_{2}\ \bar\psi(\xi_1) 
B_{2}(\xi_{1},\xi_{2})\psi(\xi_2) \ . 
\eqno({\rm III.8})
$$
where $B_{1}$ and $B_{2}(p)$ are the amplitudes for the corresponding graphs
in Fig. 1.
The tadpole amplitude $B_{1}$ is 
$$
B_{1} =  c_{1} \int dq 
{f_{j}(q)\over i q_{0}-e({\bf q})}
\eqno({\rm III.9})
$$

The Fourier transform  $\hat B_{2}(p)$ of the kernel $B_{2}$ is exactly 
$$
\hat B_{2}(p) = c_{2} \int dq_{1} dq_{2} 
{f_{j}(p+q_{1})\over i(p+q_{1})_{0}-e({\bf p+q_{1})}}
{f_{j}(-q_{1}+q_{2})\over i(-q_{1}+q_{2})_{0}-e(-{\bf q_{1}+q_{2})}}
{f_{j}(q_{2})\over i(q_{2})_{0}-e({\bf q_{2})}}
\eqno({\rm III.10})
$$
($c_{1}$ and $c_{2}$ are some inessential numerical
constant).

\noindent{\bf Lemma III.2}{ \it \ \ 
For $p$ in the support of $C^{j}$, there exists some numerical constant
$K$ such that
$$
|B_{1} |\le K  M^{2j}  
\eqno({\rm III.11})
$$
$$
|\hat B_{2}(p)| \le K  M^{2j}  
\eqno({\rm III.12})
$$
}
\prf (III.11) follows from (II.5b).
To bound $B_{2}$ we can bound the three denominators in (III.10)
by $M^{-3j}$, the two integrals on $dq_{1,0}$ and $dq_{2,0}$ 
by $M^{2j}$, and the two spatial
integrals on $d{\bf q}_{1}$ and $d{\bf q}_{2}$
by $M^{3j}$. Indeed let $S_{a}$ be the sphere of radius 1 centered
around $a$. The ${\bf q}_{1}$ integral is restricted to distance $M^{j}$
of $S_{-\p}$ by the function $e(\p + \q_{1})$, hence runs on a volume of size
$M^{j}$. But for fixed $\q_{1}$ generic, i.e.
such that $|\q_{1}|$ 
is not close to 0 or $2 \sqrt{2m\mu}$, the ${\bf q}_{2}$ integral 
is restricted to distance $\simeq M^{j}$ of the 
circle $S_{0}\cap S_{-\q_{1}}$, hence runs  
on a volume of size $O(1) M^{2j}$. The degenerate cases $|\q_{1}|
\simeq 2 \sqrt{2m\mu}$ leads also to the same bound.
\endproof

Remark also that $\hat B_{2}(p)$ is spatially rotation invariant so that
it can be written solely as a function of $p_{0}$ and $u$:
$$
\hat B_{2}(p) = \hat {\bf B}_{2}(p_{0}, u) 
\eqno({\rm III.13})
$$
with $u= e(\p)$.

By lemma III.2, the power series $\bar C^{j}(p) = \sum_{n=0}^{\infty}
C^{j}(p)\biggl((\la B_{1}+\la^{2}B_{2}(p))C^{j}(p)\biggr)^{n}$
are convergent, uniformly in $j$ and $p$, for $\la \le 0(1)$\footnote*{
By a fixed point analysis we could iterate this argument so that 
$B_{1}$ and $B_{2}$ themselves are replaced by 
similar quantities called $\bar B_{1}$and $\bar B_{2}$, computed
with propagators $\bar C^{j}$ 
instead of $C^{j}$(this is the so-called problem of ``bubbles
within bubbles''). This is not really necessary for our construction,
but would simplify the cancellation of the graphs
$B_{1}$ and $B_{2}$ with the counterterms of ${\cal M}$.}.

The propagator 
$\bar C^{j}$ is given by 
an integral representation identical to (II.1) 
but with $D(p_{0},u)= (ip_{0}-u)^{-1}$ replaced 
by $\bar D(p_{0},u)
= \bigl(ip_{0}-u - f_{j}(p_{0},u)(\la \bar B_{1}+\la^{2}
\hat {\bf B}_{2}(p_{0},u))\bigr)^{-1}$
We can slice the propagator 
$\bar C^{j}=\sum_{k}\bar C^{j,k}$ exactly as in (II.3).
Lemma 1 then applies exactly in the same way, since $\bar D$
and its derivatives satisfy the same bounds as $D$ on the support of 
$f_{j}$. This proves:

\medskip
\noindent{\bf Lemma III.3}{\it \ \ 
The sliced propagator $\bar C^{j,k}$ satisfies
the same bounds (II.5a-b) as $C^{j,k}$.}

From now on we will forget the bars, and for simplicity we
write $C^{j}$, $C^{j,k}$, $D$ instead of $\bar C^{j}$
$\bar C^{j,k}$, $\bar D$.

To obtain a better radius of convergence for our 
single slice model, we return to
the finite volume theory. For this theory,
the interaction ${\cal I}_\Lambda$ is
equal to ${\cal I}$ restricted to $\La$, hence is:
$$ 
{\cal I}_{\La}= {\cal V}_{\La} + {\cal M}_{\la} + {\cal W}_{\La}
$$
$$ {\cal V}_{\La} =
{\la\over 2}
\int_{\La^{4}} \prod_{i=1}^4 d\xi_i\ \  
V\!\left(\xi_1,\xi_2,\xi_3,\xi_4\right)
\bar\psi(\xi_1)\bar\psi(\xi_2)\psi(\xi_4)\psi(\xi_3)
\eqno({\rm III.14})
$$
$$  {\cal M}_{\La}=\la \int_{\La}d\xi B_{1} \bar\psi(\xi)\ps (\xi) + 
\int_{\La^{2}} \prod_{i=1}^2 d\xi_i\ \bar\psi(\xi_1) 
B_{2}(\xi_{1},\xi_{2})\psi(\xi_2)  
\eqno({\rm III.15})
$$
$$ {\cal W}_{\La} = |\La| ( a\la + b \la^{2} + c\la^{3} )
\eqno({\rm III.16})
$$
where ${\cal W}_{\La}$ stands for the finite volume vacuum counterterms,
and $ {\cal M}_{\la}$ for the  finite volume
two point function counterterms.

Remark that the finite volume Green's functions for that theory
this time differ by a (small)
change in the boundary condition from those of
the initial theory put in a finite volume. 
This change in the boundary condition comes from the  
fact that the counterterms
are limited to the volume $\La$, whether the power series for the
propagator is computed as if they where introduced in 
the whole space $\bbbr^{d+1}$.
But by lemma I.1 we know that we
are in a single phase, at least for $\la$ small enough, and that
the thermodynamic limit is independent
of boundary conditions. Therefore we can compute it in this way as well. 

The theory is analyzed as usual with a sequence of cluster
expansions. Let $\cD_{j}$ be a covering of $\bbbr^{d+1}$ by hypercubes of
side size $M^{-j}$, and $\cD_{k}$ be a covering of $\bbbr^{d+1}$
by hyperrectangles which have size $M^{-k}$ in each spatial direction
and size $M^{-j}$ in the (imaginary) time direction.

The finite volume $\La$
is chosen to be a finite union of $N(\La)$ hypercubes
in  $\cD_{j}$. This union is called $\cD_{j}(\La)$.
We choose $M$ integer so that each   
such hypercube is then decomposed into $M^{-3(j-k)}$ hyperrectangles of
$\cD_{k}$. Therefore $\La$ is also a union $\cD_{k}(\La)$  of hyperrectangles 
of $\cD_{k}$.

For $\De \in \cD_{k}$ we call $\ch_{\De} $ the characteristic function of
$\De$. We decompose the external functional derivatives in $\cal G$ 
according to the cubes of $\cD_{j}(\La)$ in which they act (recall that
one of them is fixed, say at the origin, and the others have to be summed
in $\La$). For any subset $X$ of hyperrectangles at higher scales
we can also define ${\cal G}_{X}$ as
those derivatives but restricted to the support of $X$.
We perform the first horizontal cluster expansion with respect 
to the $k=0$ scale. Let us denote by $<>_{\le k}$, for $k=0,...,j$,
the (normalized) Grassmann integral with respect to the Gaussian 
(anticommuting) measure of propagator
$\sum_{l\le k} C^{j,l}$.

We  consider a given Green's function
$$
\<G_{p}\>_{\le 0}={1\over Z_\Lambda}\int G_{p} 
e^{-{\cal I}_\Lambda(\psi,\bar\psi)}
d\mu_{C^{(j)}}(\psi,\bar\psi) \ ,
\eqno({\rm III.17})$$ 
where $Z_\Lambda = \int 
e^{-\lambda{\cal I}_\Lambda(\psi,\bar\psi)}
d\mu_{C^{(j)}}(\psi,\bar\psi)$ and
$$ G_{p}=\int
\prod_{j=1}^{2p} d\xi_j\ G(\xi_1,\cdots,\xi_{2p})
\prod_{j=1}^{2p}\optbar\psi(\xi_j)
\eqno({\rm III.18})
$$
is an arbitrary monomial.

 The non-local counterterm $B_{2}$ is decomposed as
$$
B_{2}(\xi, \bar \xi)= c\biggl(\sum_{k} C^{j,k}(\xi, \bar \xi)\biggr)^{3}
\eqno({\rm III.19})
$$
A standard cluster expansion with respect to 
$C^{j, 0}$ is now performed, including the propagators  
in the non-local counterterm $B_{2}$.
The formula is ([B][AR])~:

$$
\<{\cal G}\>_{\le 0} = \<  \sum_{ X_{1}, \cdots , X_{q} \atop X_{1},\cdots ,
X_{q} \ {\rm disjoint, \ non-trivial} } {1 \over q~!}
\prod_{i=1}^{q}
\cA (X_{i})\>_{\le -1}
\eqno({\rm III.20})$$
The polymers $X_{i}$ are non-empty sets of hyperrectangles of $\cD_{0}$, and
the amplitude $\cA (X_{i})$  
for the polymer $X_{i}$ is given by a ``Brydges formula''
$$
\cA (X) = {1 \over Z_0} \sum_{T}  \int d^{T}h \int d\mu_{C^{j,0}(T,h)}
\biggl\{
\prod_{l \in T}   L_{l} \biggr\} 
{\cal G}_{X}e^{-\lambda{\cal I}_X(\psi,\bar\psi)}  
\eqno({\rm III.21})
$$
where $Z_0$ is the normalization factor of a trivial
polymer made of a single hyperrectangle with no external legs.
The sum is performed
over oriented trees built on the polymer $X$. The orientation tells us
when a tree line $l$ joins $X$ to $X'$ if it is 
$\xi$ which is located in $X$ and $\bar \xi$ in $X'$,
in which case we put $\De_{l}=\De$ and $\bar \De_{l}=\De '$, 
or the contrary, in which case we put $\bar\De_{l}=\De$ and $\De_{l}=\De '$.
The linking operators are
$$
L_{l} = \int_{\De_{l}} d\xi  \int_{\bar \De_{l}}
d\bar \xi C^{j,0}(\xi, \bar \xi) {\de
\over \de \psi_{0} (\xi)} {\de
\over \de \bar \psi_{0} (\bar \xi)} $$
$$+ 3c
\int_{\De_{l}} d\xi  \int_{\bar \De_{l}}
d\bar \xi \bar\psi(\bar\xi) 
\psi(\xi)C^{j,0}(\xi, \bar \xi)\biggl(
C^{j,0}(T,h,\xi, \bar \xi)+\sum_{k\le -1} C^{j,k}(\xi, \bar \xi)
\biggr)^{2}
\eqno({\rm III.22})
$$
The second term corresponds to the non-local coupling through the
$B_{2}$ counterterm.

$d^{T}h$ is the ordinary Lebesgue measure over interpolation parameters
$h=\{h_{l}, h\in T\} $, each running in $[0,1]$. 
$d\mu_{C^{j,0}(T,h)}$ is the normalized Grassmann Gaussian measure 
with propagator $C^{j,0}(T,h)$, depending on the interpolation
parameters $h$, which is equal to $C^{j,0}(\xi, \bar \xi) \cdot 
h^{T}(\xi, \bar \xi)$, where $h^{T}(\xi, \bar \xi)$ is 0 if 
$\xi$ or $\bar \xi$ are not in the support $X$ of the polymer,
is 1 if they lie in the same hyperrectangle of $X$, and is otherwise the infimum
over all parameters $h_{l}$ on the unique path joining the hyperrectangle containing
$\xi$ to the hyperrectangle containing $\bar\xi$ [B][AR].

Then we apply a
vertical expansion which tests whether a polymer $X$
contains or not fields of lower scales (recall that external fields in $\cG$
are considered by convention of the scale $j-1$, 
so lower than all other fields).
This means that we introduce a parameter $v_{X}$ which for each $X$
interpolates:

$$
{\cal V}_{X}={\cal V}_{X}^{0}+{\cal V}_{X}^{\le -1}+ 
{\cal V}_{X}^{0{\rm \ linked\  to\ }\le-1}
\eqno({\rm III.23})
$$
$$
{\cal V}_{X}^{0}={1\over 2}\int_{X} 
d \x d \ta \bar\psi^{0}_{\uparrow}(\x, \ta)\bar\psi^{0}_{\downarrow}
(\x, \ta)\psi^{0}_{\uparrow}(\x, \ta)\psi^{0}_{\downarrow}(\x, \ta)
\eqno({\rm III.24a})$$
$$
{\cal V}_{X}^{\le -1}
= {1\over 2}\int_{X}\bar\psi^{\le -1}_{\uparrow}(\x, \ta)
\bar\psi^{\le -1}_{\downarrow}
(\x, \ta)\psi^{\le -1}_{\uparrow}(\x, \ta)\psi^{\le -1}_{\downarrow}(\x, \ta)
\eqno({\rm III.24b})$$
$$
{\cal V}_{X}^{0{\rm \ linked\  to\ }\le-1}=
{1\over 2}\int_{X} d \x d \ta \bar\psi_{\uparrow}(\x, \ta)\bar\psi_{\downarrow}
(\x, \ta)\psi_{\uparrow}(\x, \ta)\psi_{\downarrow}(\x, \ta)
$$
$$-\int_{X} d \x d \ta \bar\psi^{0}_{\uparrow}(\x, \ta)
\bar\psi^{0}_{\downarrow}
(\x, \ta)\psi^{0}_{\uparrow}(\x, \ta)\psi^{0}_{\downarrow}(\x, \ta) -
\bar\psi^{\le -1}_{\uparrow}(\x, \ta)\bar\psi^{\le -1}_{\downarrow}
(\x, \ta)\psi^{\le -1}_{\uparrow}(\x, \ta)\psi^{\le -1}_{\downarrow}(\x, \ta)
\eqno({\rm III.24c})$$
$${\cal V}_{X}(v_{X})={\cal V}_{X}^{0}+{\cal V}_{X}^{\le -1}+ 
v_{X}{\cal V}_{X}^{0{\rm \ linked\  to\ }\le-1} \ ; \ 
{\cal V}_{X} =    {\cal V}_{X}(v_{X})|_{v_{X}=1}
\eqno({\rm III.25})
$$
and we also interpolate ${\cal M}_{X}$ in a similar way.
Then we perform a Taylor expansion in $v_{X}$ to first order for each $X$.
The terms with $v_{X}=0$
are called  vacuum polymers. For the error terms
we draw a link between the polymer $X$ and  
all the hyperrectangles of the next scale containing a hyperrectangle
of its support; in this way we construct a polymer $\tilde X$
living on the hyperrectangles of the next scale.

Then we apply a contraction rule (integration by parts on the field), 
to compensate  explicitly 
the vacuum polymers which contain
up to three vertices with the corresponding counterterms,
and the polymers with exactly one or two vertices and two low momentum fields
of scale $\le -1$. This is standard [GJ]. There remains only polymers 
containing at least 4 vertices or at least 4 external legs 
(a $B_{2}$ counterterm counts for 2 vertices, etc...).

This explicit rule
a la Glimm-Jaffe is slightly simpler than
the modern more powerful renormalization
group schemes that are needed for just renormalizable theories [B],[R].
Indeed it avoids to perform at each scale a Mayer 
expansion to factorize {\it all} the vacuum graphs,
and e.g. exponentiate {\it all} the two point functions. 
The result of a   Mayer expansion 
is a formula expressed in terms of ``Mayer configurations'',
i.e. finite sequences of polymers. Iteration of Mayer expansions lead 
to sequences of sequences, hence to formulas
heavier to manipulate than ordinary polymers.

To iterate we apply the second scale cluster expansion 
to the set of non trivial subsets of $\cD_{-1}$
whose elements are the polymers 
obtained after the first expansion,
and we apply a second cluster expansion with respect to the measure
$C^{j, \, -1}$ between these units, etc...

We obtain the following final formula

$$
\<{\cal G}\>_{\le 0} = \sum_{ Y_{1}, \cdots , Y_{q} 
\atop Y^{k}_{1}(Y),\cdots ,
Y^{k}_{q}(Y) \ {\rm disjoint}}
{1 \over q!} \prod_{i=1}^{q}
\cA (Y_{i})
\eqno({\rm III.27})$$
The polymers $Y$ at the last scale have vertical tree structure,
that is they are a collection of $1+|j|$ sets $Y^{k}$ 
for $k=0,\cdots,j$. More precisely,
$Y^{k}(Y)$ is a subset of $\cD_{k}$, i.e. a set of hyperrectangles, 
made of connected components $Y^{k}_{l}$ (taking at level $k$
into account the connections of scales higher than or equal to $k$,
and the set of all the $Y^{k}_{l}$ has tree
structure for the inclusion relation (see e.g. [R]).   

The amplitude $\cA (Y_{i})$  
for the polymer $Y_{i}$ is given by applying inductively the Brydges
formula, according to the collection of subsets $X^{k}(Y)$.

$$
\cA (Y) = \sum_{T_{0},...,T_{j}} ... \int d^{T}s \int d\mu_{C^{j}_{0}(s)}
\biggl\{
\prod_{l \in T} L'_{l}\biggr\} 
{\cal G}_{X}e^{-\lambda{\cal I}_Y(\psi,\bar\psi)}  
\eqno({\rm III.28})
$$
where 
the links operators $L'_{l}$
include the explicit contraction rules to cancel
the counterterms for 
the graphs $B_{1}$, $B_{2}$
and $G_{1}$,..., $G_{3}$
and the sum in (III.28) is restricted by the condition that any
connected component at a given level has either more than three
vertices or more than two external legs.

Theorem I then follows from the following bound

\medskip
\noindent{\bf Theorem III.1}{\it\ \  For any $K>0$
there exists some $r(K) >0$ such that for 
$|\la|\le r(K)$
$$
\sum_{Y | 0 \in X^{0}(Y)} | \cA (Y) | K^{|X^{0}(Y)|} \le 1
\eqno({\rm III.29})
$$}
Indeed performing a standard Mayer expansion we obtain analyticity 
of $<\cG>$
for $|\la|\le r(e)$ (where $e=2.718..$).

The next section is devoted to a proof of Theorem III.1.

\medskip
\noindent{\bf IV The bounds}

The formula giving the amplitude of a polymer $Y$ is expanded into a sum
of diagrams made of explicit propagators derived by the cluster expansion, 
vertices, and 
fields. The Fermionic Gaussian integration on the fields 
is simply a formal device which changes the fields into
products of determinants.
We recall the ``method of combinatorial factors'' to keep track of the
many sums in the expansion. This technique uses the elementary
estimate

$$
\kappa_i>0\ ,\ \sum_i \kappa_i^{-1}\le 1\ \ \     
\Rightarrow\ \ \ \left|\sum_i U_i\right|\le\sup_i
\left|\kappa_i U_i\right|.
\eqno({\rm IV.1})$$   
to replace each sum by a supremum. To help remember the combinatorial
factor $\kappa_i$ multiplying the value $U_i$ of a given diagram, the
factor will be assigned to a specific line or vertex 
or field of the diagram. Remember also that in the end 
we have a small coupling constant per vertex, hence per field,
so that such constants can be forgotten in the estimates.

A vertex will be localized in the tube of its highest leg.
That means that we write 
$$
e^{{\cal V}_{Y}} = e ^{\sum_{\De \in Y}  \int _{\De} d\xi
\optbar\ps ^{k(\De)}(\xi)\sum_{k_{1}, k_{2}, k_{3}\le k(\De)} 
\optbar\ps ^{k_{1}}(\xi) \optbar\ps ^{k_{2}}(\xi) \optbar\ps ^{k_{3}}(\xi)  }
$$
$$  =\prod_{\De} \sum_{v(\De)=0}^{\infty}{1\over v(\De)!}  \biggl( \int _{\De} 
d\xi
\optbar\ps ^{k(\De)}(\xi)\sum_{k_{1}, k_{2}, k_{3}\le k(\De)} 
\optbar\ps ^{k_{1}}(\xi) \optbar\ps ^{k_{2}}(\xi) \optbar\ps ^{k_{3}}(\xi) 
\biggr)\eqno({\rm IV.2}) 
$$
Furthermore by the usual ``local factorial principle'' we can extract, up to a
constant per cube of $Y$,
from the decay of the explicit cluster propagators a large power
of the product for all tubes of the factorial of 
the number of the vertices produced 
by the explicit cluster functional derivations and integration by parts [R].
This is useful because we have no symmetry factor in ${1\over v(\De)!}$
for these vertices.
Joining this fact to  (IV.2), we have a factor ${1\over v(\De)!}$
where $v(\De)$ is now the total number of vertices in the polymer
localized in $\De$, no matter whether they are hooked to explicit
cluster derivatives or not.

For any field $\optbar\ps^{k}$, we call $k(\optbar\ps)$ its scale $k$
and  $l(\optbar\ps)$ the scale of the localization tube of the vertex to which
it is hooked. We call $\De_{l}(\optbar\ps)$ the localization tube
of the vertex to which it is hooked, not to be confused
with the tube $\De(\optbar\ps)$ of scale $k(\optbar\ps)$ in which it lies.
Remark that $\De_{l}(\optbar\ps)\subset \De(\optbar\ps)$.

By a  combinatoric factor $\ka_{1}=M^{(\ep/4)(k(\optbar\ps)-j)}$ we 
can sum over all
scales of all legs of any vertex. In what follows $\ep$ is a very small number.

By a  combinatoric 
factor $\ka_{2}=M^{-j -3 k(v)}$ per vertex of localization scale $k(v)$
we can integrate over the position of that vertex in its localization cube.

Then at each level $k$ 
let us consider the piece $Y_{k}$ made of the tubes of $Y$
of scale $k$. Each tube contains a certain number $f(\De)$ of fields
(or anti fields) of scale $k$ lying in this cube.

We decompose the covariance $C^{k}(h,T)$ 
as $C^{k}(h,T) $ $=$ 
$ \sum_{\De, \De' \in X_{k}}  \ch_{\De}C^{k}(h,T)\ch_{\De'}$
and correspondingly we expand the fields.
Let us introduce the scaled distance between 
$\De$ and $\De'$ as 
$$d(\De, \De') = \inf_{\xi \in \De, \bar\xi \in \De'}
(M^{k}|\x -\y|+M^{j}|\tau -\bar \tau|) 
\eqno({\rm IV.3})$$
By a combinatoric factor $\ka_{3}=(1+d(\De, \De'))^{5} $ 
we can choose, for each
field or anti field, the tube (of the same scale) to which it contracts.

Let $\optbar f(\De,\De')$ be the number of fields (respectively antifields)
localized in $\De$ that contract to a field localized in $\De'$.
By Fermionic rules of integration, $f(\De, \De')=\bar f(\De', \De)$.
We obtain therefore after Gaussian integration 
for each scale a product of determinants

$$
\prod_{(\De,\De')\in Y_{k}^{2}} \det\nolimits_{(\De,\De')} 
\eqno({\rm IV.4})$$
where $\det_{(\De,\De')} $ is a $f(\De, \De')$ by $\bar f(\De', \De)=
f(\De, \De')$ determinant of propagators $C^{j,k}$. Applying
the Hadamard's bound $|\det a_{ij} |\le n^{n/2}\sup |a_{ij}|^{n} $
we obtain by (II.5a) for these determinants
a factor $f(\De, \De')^{f(\De, \De')/2}  
(1+d(\De, \De'))^{-p f(\De, \De') }  M^{(j+k)f(\De, \De')}$.
Taking $p$ larger than 10, the factor $(1+d(\De, \De'))^{-p/2}$ per field
absorbs the combinatoric factor $\ka_{3}$.
Joining together the lines power counting
$\prod_{k}\prod_{(\De,\De')\in Y_{k}^{2}} 
M^{(j+k)f(\De, \De')}$ given by the determinant
and by the explicit propagators to the integration factor 
$\ka_{2}$ we obtain the power counting factor
$M^{(j-k(v)) - \half\sum_{i=1,2,3}(k(v)-k_{i})}$ per vertex,
which is bounded by a factor 
$M^{(j-k(v))(1/4-\ep)}$ per vertex times a factor 
$M^{((3/4)+\ep)(k(\optbar\ps)-l(\optbar\ps))}$ per field.
Therefore combining this factor to the factor $\prod_{\De}{1\over v(\De)!}$
we have a factor
$$
M^{((3/4)+\ep)(k(\optbar\ps)-l(\optbar\ps))}{v(\De_{l}(\optbar\ps))^{-1/4}} 
\eqno({\rm IV.5})
$$
per field. After having absorbed the combinatoric factor $\ka_{1}$
we also have still a factor
$M^{(j-k(v))(1/4-2\ep)}$ per vertex.

Now 
we bound the factorials of Hadamard's inequality
using in particular  $f(\De, \De')=\bar f(\De', \De)$:

$$
\prod_{k}\prod_{(\De,\De')\in Y_{k}^{2}}f(\De, \De')^{f(\De, \De')/2}
\le \prod_{k}\prod_{(\De,\De')\in Y_{k}^{2}}f(\De, \De')
^{f(\De, \De')/4}
\bar f(\De, \De')^{\bar f(\De, \De')/4}
$$
$$\le \prod_{k}\prod_{(\De,\De')\in Y_{k}^{2}} 
(f(\De, \De')+\bar f(\De, \De'))
^{f(\De, \De')/4+\bar f(\De, \De')/4}
$$
$$\le \prod_{k}\prod_{\De\in Y_{k}} f(\De)
^{\sum_{\De'}f(\De, \De')/4+\bar f(\De, \De')/4}
$$
$$\le \prod_{k}\prod_{\De\in Y_{k}} f(\De)
^{f(\De)/4} \le  \prod_{k}\prod_{\De\in Y_{k}} K^{f(\De)} f(\De)!!^{1/2}
\eqno({\rm IV.6})
$$
So up to an inessential constant per field, Hadamard's bound
gives a product over all tubes of the sqare root of
the number of all Wick contractions
between all fields localized in
this tube, not taking into account whether they are fields or anti-fields.
(Remark that $f(\De)$ can be odd, in which case we put by convention
$f(\De)!!= (f(\De)-2)...5.3.1$, hence we do not contract one of the legs).

Then we have:

\noindent{\bf Lemma IV.1}
{\it
$$ \prod_{k}\prod_{\De\in Y_{k}} ( f(\De)!!)^{1/2}
\le \prod_{k}\prod_{\De\in Y_{k}} K^{f(\De)}
\prod_{\ps \in \De} \biggl( 
M^{(3+\ep/4)(l(\optbar\ps)-k(\optbar\ps)) } v(\De_{l}(\ps))\biggr)^{1/4}
\eqno({\rm IV.7})
$$
}
\prf We can contract the legs starting with the ones which have largest
value of the factor
$M^{(3+\ep/4)(l(\optbar\ps)-k(\optbar\ps)) } v(\De_{l}(\ps))$,
and with a factor 
$$M^{(3+\ep/4)(l(\optbar\ps ')-k(\optbar\ps ')) } 4v(\De_{l}(\ps ')) 
$$
$$\le 
\sqrt{M^{(3+\ep/4)(l(\optbar\ps )-k(\optbar\ps )) } 4v(\De_{l}(\ps ))
M^{(3+\ep/4)(l(\optbar\ps ')-k(\optbar\ps ')) } 4v(\De_{l}(\ps '))}
\eqno({\rm IV.8})$$
we can choose to which leg it contracts, by choosing first 
with $M^{(3+\ep/4)(l(\optbar\ps ')-k(\optbar\ps ')) }$
the localization tube of the leg to which it contracts (this is because
there are $M^{3(k-k')}$ tubes of scale $k$ in a tube of scale $k'\le k$), 
then by a factor
$v(\De_{l}(\ps '))$ the vertex in this localization tube, then with a factor
4 the particular leg. Collecting these factors proves the lemma. \endproof

Comparing the factor of (IV.7) to the one of (IV.5)
we see that we get a constant per field.

It remains to sum over the positions of the tubes in $Y$.
This is standard
ordinary power counting and it can be done with the 
decay of the cluster propagators in the horizontal direction
(at fixed $k$) plus the power counting 
of the vertices involved in the vertical
connections, 
using the fact that no connected subgraph
$Y_{k}^{l}$ has less than two external legs and less than three vertices.

Finally the constant per field is controlled by the smallness
of $\la$. This completes the proof of Theorem III.1. Taking into account the
global fixed factor $M^{j(4-2p)}$ due to our norm, Theorem I follows.

\medskip
\noindent{\bf  Appendix I}
\medskip
In this appendix we explain the slight modifications to adapt the proof 
of Theorem I to all space dimensions greater than three.

The Fourier transform of the $\de$ function of the sphere $S^{d-1}$ in 
$\bbbr^{d}$ behaves as $1/|x|^{(d-1)/2}$ at large $x$.
Therefore reslicing the corresponding propagator $C^{j}$ in auxiliary
slices $C^{j,k}$ we have 
$$
\big| 
C^{j,k}(\xi,\bar \xi) \big| 
\le K(p) M^{j+ k(d-1)/2} 
(1+M^{k}|\x -\y|)^{-p} (1+M^{j}|\tau -\bar \tau|)^{-p} 
\eqno({\rm A.I.1})
$$

The power counting of a graph
with $n$ vertices at scale $k$ and $e=2p$ external legs at scale $j$ is 
$M^{(d+1-p(d+1)/2)j} M^{(j-k)(n+p(d-1)/2-d)}$. This is obtained by combining
the power counting $M^{(j+k(d-1)/2)(2n-p)}$
of the legs at scale $k$, and  the power counting $M^{-(j+dk)(n-1)}$
of the $n-1$ vertices integration. Therefore the auxiliary problem
is superrenormalizable in any dimension. There is only a finite number
of graphs to renormalize, those satisfying $n+p(d-1)/2 \le d$,
which are always vacuum ($p=0$) or
two point ($p=1$) subgraphs.  
The main result of this paper, Theorem I
extends therefore without difficulty
to any dimension, with the prefactor $M^{(4-2p)j}$
in (I.15) replaced by $M^{(d+1-p(d+1)/2)j}$.

It is interesting to remark that as $d\to \infty$
more and more 2-point subgraphs diverge, and we obtain real
Fermionic models (in integer dimensions) 
whose power counting mimick the ill-defined
$\ph^{4}$ models in fractional dimensions $d\to$ 4. 
The ``critical'' dimension at which
our problem becomes just renormalizable is 
$d=\infty$, and it would presumably be interesting to set up for it
a $1/d$ expansion which would be the analogue
of the Fisher-Wilson $1/\ep$ expansion for $\ph^{4}$ models. 

\medskip
\noindent{\bf Appendix II}
\medskip

In this appendix we include a discussion of why the sector
decomposition used in [FMRT1] in the two-dimensional case does not
work in $d=3$. Since the volume of the $j$-th momentum
shell is large, it is natural 
to decompose ths shell into $M^{-(d-1)j}$ sectors of side $M^{j}$,
to get fields which are supported on cells of unit volume in phase space.
This canbe done through a smooth partition of unity
$$
1=\sum_{m=1}^{M^{-(d-1)j}}\eta_m(\p),\hskip .25truein
\eta_m(\p)=\eta_m\left({\p\over|\p|}k_F\right)
$$
of the Fermi surface, where $\eta_m$ is supported on the union of the
$m^{th}$ sector, $S_m$, and its neighbors, whose number is at most
$3^{d-1}-1$. There is a corresponding decomposition of the covariance
$$
C^{(j)}=\sum_{m=1}^{M^{-(d-1)j}} C^{(j,m)}
$$ 
where
$$
C^{(j,m)}(\xi,\bar \xi)=\delta_{\sigma,\bar\sigma}
\int {d^{d+1}p\over(2\pi)^{d+1}} {e^{i<p,\xi-\bar\xi>_-}
\over ip_0-e(\p)}f_j(p)\,\eta_m(\p)
$$
and of the fields
$$
\psi^{j}=\sum_{m=1}^{M^{-(d-1)j}}\psi^{(j,m)},\hskip.25truein
\bar\psi^{j}=\sum_{m=1}^{M^{-(d-1)j}}\bar\psi^{(j,m)}.
$$

The standard power counting bound for an individual graph is still
easy when there are sectors. First, one selects a spanning
tree for the graph. To each line not in the tree there is a
corresponding momentum loop, obtained by joining its ends through a
path in the tree. This construction produces a complete set of
independent loops. Ignoring unimportant constants, each propagator is
bounded by its supremum $M^{-j}$.  The volume of integration for each
loop is now $M^{(d+1)j}$. A priori, there is one sector sum with
$M^{-(d-1)j}$ terms for each line. But, by conservation of momentum,
there is only one sector sum per loop. Thus, if there are $n$ vertices
and $e=2p$ external lines, the supremum in momentum space of the graph is
bounded by $$\eqalign{
\prod_{\rm lines}M^{-j}\prod_{\rm loops}M^{(d+1)j}M^{-(d-1)j}
&=M^{-j(4n-2p)/2}M^{2j\left[(4n-2p)/2-(n-1)\right]}\cr &=M^{{1\over
2}(4-2p)j}.  
}$$

In the course of a non-perturbative construction, estimates cannot be
made graph by graph because there are too many of them. Rather, the
perturbation series must be blocked and the blocks estimated as units.
The blocks are estimated using the exclusion principle to implement
strong cancellations between the roughly $n!^2$ graphs of order $n$.
However, once the series is blocked, momentum loops can't be defined
and the argument leading to the estimate above cannot be made.
Conservation of momentum has to be implemented at each vertex, rather
than through loops. Even though the volume cutoff $\Lambda$ breaks
exact conservation of momentum, many of the $M^{-2\ell(d-1)j}$ terms
in the sector sums for a general $2\ell$-legged vertex must be zero.

In [FMRT1] the following lemma is proved:

\noindent{\bf Lemma 2}
{\it  Fix $\ m\in {\bf Z}^{d+1}\ $ and $\ \ell\ \ge\ 2\ $.  
Then, the number of $2\ell$-tuples 
$$
\left\{S_1,\ \cdots\,S_{2\ell}\right\}
$$ 
of sectors for which there exist $\ \k_i\in \bbbr^d,\ i=1,\ \cdots,\ 
2\ell\ $ satisfying 
$$
\k_i^\prime\in S_i,\ \  |\k_i-\k_i^\prime|\le {\rm const}\ M^j
\ , \ \ \ i=1,\ \cdots,\ 2\ell\ 
$$ 
and 
$$
 |\k_1+\ \cdots\ +\k_{2\ell}|\le{\rm const}\left(1+|m|\right)M^j 
$$ 
is bounded by 
$$
\const^\ell(1+|m|)^dM^{-(2\ell-1)(d-1)j}M^j
\left\{1+|j|\delta_{d,2}\delta_{\ell,2}\right\}.
$$ 
In particular, for a four legged vertex, the number of $4$-tuples is
at most 
$$
\const(1+|n|)^dM^{(-3d+4)j}\left\{1+|j|\delta_{d,2}\right\}.
$$ 
Here, $\ \k^\prime = {\k\over |\k|}\ $ denotes the projection of $\
\k\ $ onto the Fermi surface.}
\vskip.25truein

This lemma although instructive, is neither for $d=2$ nor for $d>2$
powerful enough for the non-perturbative construction. However for
$d=2$ the number of active sector 4-tuples at a vertex is of
order $\ |j| M^{-2j}\ $; ordinary power counting at a vertex
can accomodate at most $M^{-2j}$, hence there is only a ``logarithmic''
power counting deficit of order $|j|$ at each vertex. For $d=3$
the number of active sector 4-tuples at a vertex is oforder $\ M^{-5j}\ $, 
and ordinary power counting at a vertex
can accomodate only $\ M^{-4j}\ $ so that the deficit 
at each vertex is a full power $M^{-j}$. This deficit is worse and
worse as $d$ increases beyond 3. In three dimension
this difficulty can be geometrically thought
as the twist or torsion which occurs in the figure
made by four momenta of unit length adding up to 0 (this figure is no longer
a planar rhombus as in two dimensions).
 
To circumvent the extra logarithm at $d=2$, in [FMRT1] we divided the Fermi
surface into sectors of length $M^{j/2}$ rather than $M^j$. Taking
into account the anisotropic spatial decay associated to these new
``rectangular'' sectors beats the logarithmic deficit. If we apply the
same idea for $d=3$, the number
of active  4-tuples sectors at a vertex becomes of order $\ M^{-5j/2}\ $
but the power counting in the manner of [FMRT1] gives $\ M^{-2j}\ $,
so that there still remains a  deficit of order $M^{-j/2}$ per vertex.
Making the sectors still longer does not work because the anisotropic
decay no longer holds since the Fermi surface does not look flat for
lengths bigger than $M^{-j/2}$. 
Remark that this difficulty
is presumably linked to the graph $G_{2}$, which as we recall is in
$M^{-j} \la^{2}$ in a cube of size $M^{-j}$.

The conclusion is that even the single slice theory in $d \ge 3$
is a non-trivial theory that contains a rather non-trivial 
renormalization, like $\ph^{4}_{3}$. It is this renormalization
that can be analyzed by means of the auxiliary
scales used in this paper.

\medskip
\noindent{\bf Appendix III}
\medskip

In this appendix we discuss briefly the present situation for the
full (multi-scale) theory.

The most straightforward 
strategy in the case of multi-scale Fermionic models
consists in writing the effect of inductive integration
of higher scales in the form of an effective action for the single scale 
model, and to check some inductive estimate for this effective action.
This is the road followed in [FMRT1]. However when
combined with the Hadamard estimate 
on determinants, this approach runs
into a major difficulty: even with a quartic bare action
the effective action contains irrelevant
operators of degree 6 and more in the fermionic fields. 
The Hadamard inequality when applied to such operators does not
lead to a convergent answer (it develops divergent factorials).

The obvious thing to do seems to treat the theory as a bosonic one.
For bosonic theories we know well
that one cannot simply exponentiate all the effective
action, as in Wilson's original program, but 
that one must keep the irrelevant effects coming
from higher scales in the form of a polymer gas with hardcore constraints
[B][R]. A typical stability estimate in this strategy is to prove that
the ``completely convergent'' multi-scale polymers 
(i.e. those who do not contain any two or four point subcontributions)
form a geometric convergent series. At the moment our best estimates
with the Hadamard method indicate that the sum over such
``completely convergent'' polymer, extending over
at most $N$ different scales, converges
provided the (bare) coupling constant $\la$ is small, $and$
the number of scales is limited to $N \le \ep / \la$
($\ep$ being a small number). This
is encouraging, but tantalizing. Indeed although this last estimate is
much better than what a sector analysis would provide 
(namely $N \le K|\log \la|$), it is still not sufficient.
For the construction of the BCS3 theory,
where the BCS gap cuts the renormalization group
flow at a scale $\De_{BCS} \simeq M^{-N} \simeq e^{-O(1)/\la}$, we would need
$N \simeq K /\la$ (where $K$ is not particularly small). For 
the construction of 
non-superconducting Fermi liquids as in [FKLT],
the renormalization group flows for ever and we would 
need $N$ independent of $\la$. Therefore
at the moment we consider that the question of the non-perturbative 
stability of these theories remains open, but we hope that the Hadamard
method, clearly much better than the sector method, is a step
towards the final solution.

\vskip 1cm
\noindent{\bf References}
\item{[AR]}{A. 
Abdesselam and V. Rivasseau, Trees, forests and jungles: a
botanical garden for cluster expansions,  
in ``Constructive Physics'', Lecture Notes in Physics 446 (Springer) (1995)


\item{[B]} D. Brydges, Weak Perturbations of the Massless
Gaussian Measure (and references therein)
in ``Constructive Physics'', Lecture Notes in Physics 446 (Springer) (1995).

\item{[FKLT]} J. Feldman, H. Kn\"orrer, D. Lehmann
and E. Trubowitz, ``Fermi liquids in Two-Space Dimensions'',
in ``Constructive Physics'', Lecture Notes in Physics 446 (Springer) (1995).

\item{[FMRT1]} J. Feldman, J. Magnen, V. Rivasseau
and E. Trubowitz, An infinite Volume Expansion for
Many Fermions Green's Functions, 
Helv. Phys. Acta, Vol. 65, 679 (1992).

\item{[FMRT2]} J. Feldman, J. Magnen, V. Rivasseau
and E. Trubowitz, Ward Identities and a Perturbative Analysis of a U(1)
Goldstone Boson in a Many Fermion System,
Helvetica Physica Acta {\bf 66}, 498 (1992).

\item{[FMRT3]} J. Feldman, J. Magnen, V. Rivasseau
and E. Trubowitz, Constructive Many Body Theory, in ``The State of Matter'',
Eds M. Aizenman and H. Araki, Advanced Series in Mathematical Physics
Vol. 20, World Scientific (1994).

\item{[FMRT4]} J. Feldman, J. Magnen, V. Rivasseau
and E. Trubowitz, An Intrinsic 1/N Expansion for Many Fermion System,
Europhys. Letters 24, 437 (1993).

\item{[FMRT5]} J. Feldman, J. Magnen, V. Rivasseau
and E. Trubowitz, Two dimensional Many Fermion Systems as Vector Models,
Europhys. Letters 24, 521 (1993).

\item{[FT1]} J. Feldman, E. Trubowitz, Perturbation Theory for Many 
Fermion Systems, Helvetica Physica Acta {\bf 63}, 156-260 (1990). 

\item{[FT2]} J. Feldman, E. Trubowitz, The Flow of an Electron-Phonon
System to the Superconducting State, Helvetica Physica Acta {\bf 64},
213-357 (1991).

\item{[GJ]} {J. Glimm and A. Jaffe, Positivity of the $\phi^{4}_{3}$  
Hamiltonian, Fortschr. Phys. 21, 327 (1973).} 

\item{[R]}{V.Rivasseau, From perturbative to constructive renormalization, 
Princeton University Press (1991).}

\end